\documentclass[11pt]{article}
\usepackage{setspace}

\def\be{\begin{equation}}

\def\ee{\end{equation}}

\def\bea{\begin{eqnarray}}

\def\eea{\end{eqnarray}}

\newcommand\eps{\epsilon}

\begin{document}

\singlespace

\begin{flushright} BRX TH-614 \\
CALT 68-2764
\end{flushright}

\vspace*{.3in}

\begin{center}

{\Large\bf   Circular Symmetry in Topologically Massive Gravity}

{\large S.\ Deser$^1$ and J.\ Franklin$^2$}

{\it $^1$Physics Department,  Brandeis University, Waltham, MA 02454 and \\
Lauritsen Laboratory, California Institute of Technology, Pasadena, CA 91125 \\
{\tt deser@brandeis.edu}

$^2$ Reed College, Portland, OR 97202, USA\\

{\tt jfrankli@reed.edu}}

\end{center}

\begin{abstract}
We re-derive, compactly, a TMG decoupling theorem: source-free TMG separates into its
Einstein and Cotton sectors for spaces with a hypersurface-orthogonal Killing vector, here concretely for circular symmetry. We can then generalize it to include matter, which is necessarily null.
 \end{abstract}

  \section{Introduction}
  
Topologically massive gravity (TMG) [1] is a counterexample to almost all standard lore. The sum of  ordinary Einstein ($G_{\mu\nu}$) and Cotton-Weyl ($C_{\mu\nu}$) sectors in $D=3$, it contains very nontrivial bulk excitations and solutions. Yet the constituent sectors are separately trivial: all Einstein solutions have locally constant curvature (or are flat if $\Lambda=0$); vanishing Cotton implies conformally flat space, including of course (A)dS. [While solutions of pure GR always trivially satisfy TMG (and Cotton), they are not, in general, its only solutions.] This raises the converse question: under what conditions will the combined system necessarily re-dissolve into its (trivial) constituents? Remarkably, a general decoupling criterion exists [2]: presence of a hypersurface-orthogonal Killing vector (HSOK) $X_\mu$. The, somewhat abstract, derivation of [2] is based on the ``kinematical" lemma that, for each possible component projection, along and orthogonal to $X_\mu$ -- just one of the respective components of the Ricci and Cotton tensors vanishes identically. Applied to source-free TMG, this implies the separate vanishing of the two sectors' tensors, reducing the solutions to those of GR -- no ``true" TMG extensions exist. Our aim here is the twofold one of tracing this decoupling to its cause -- a ``mismatch" between Einstein and Cotton tensors, thereby providing a short, simple, proof -- then to analyze its applicability in the presence of matter. For concreteness, we use the most familiar and important HSOK, circular ($D=2$) symmetry, but the results are general.



The TMG equations with a cosmological term are  
\begin{eqnarray}
E^{\mu\nu} &\equiv&  \sqrt{-g} \, (G^{\mu\nu}+ \Lambda\,  g^{\mu\nu})+ m^{-1}\, C^{\mu\nu} = \kappa\,  T^{\mu\nu},  \nonumber \\
C^{\mu\nu} &\equiv& \eps^{\mu \alpha \beta}\,  D_\alpha \, S_\beta^\nu\, ,  \hbox{\hskip 1cm} S_\beta^\nu \equiv   [R_\beta^\nu - 1/4\,  \delta_\beta^\nu\,  R].  
\end{eqnarray}
For simplicity of notation (only), the $\Lambda$-term is understood implicitly in $G$ below. Also, we set $T^{\mu\nu}=0$ 
to start with. The key to decoupling is the Levi-Civita tensor, $\eps^{0ij} \equiv \eps^{ij}$, in $C^{\mu\nu} \equiv C^{\nu\mu}$, along with the elementary fact that, in circularly symmetric (but not necessarily time-independent) spaces, all  $2$-vectors and their axial versions are proportional to $x^i$ and $\eps^{ij}\,  x^j$ respectively, and their $2$-tensor equivalents to ($x^i\,  x^j$, $\delta^{ij}$) and $\eps^{k(i}\,  x^{j)}\,  x^k$. [We will use this simple notation instead of the more abstract one in terms of $X_\mu= g_{\mu \phi}$.] An immediate consequence is that the $2$-(pseudo)scalar $C^{00}$, being proportional to $\eps^{ij}$, vanishes identically, implying $G^{00}=0$. The mixed term,  
\begin{equation}
E^{0i} =a\, x^i + m^{-1}\, b \, \eps^{ij}\,  x^j =0,  
\end{equation}
forces the two functions $(a,b)(r,t)$ to vanish separately, as is obvious by projecting (2) with $x^i$ or $\eps^{ik}\,  x^k$: this  means $G^{0i}=0=C^{0i}$. The spatial components,  
\begin{equation}
E^{ij}= [c\,  x^i \, x^j + d\,  \delta^{ij}] + m^{-1}\,  f\,  \eps^{k (i}\,  x^{j)}\,  x^k =0,      
\end{equation}
may be projected with the (even) $x^i\,  x^j$ and $\delta^{ij}$ to show that both $(c,d)(r,t)=0$, hence also $f=0$, that is $G^{ij}=0=C^{ij}$. This completes the short proof that all $G_{\mu \nu}$ and $C^{\mu \nu}$ components of source-free TMG vanish separately in presence of HSOK, due to the $\eps^{ij}$-``mismatch". 

Next, we  try to include a (circularly symmetric, of course) source, for example an imploding circular matter shell, by reinstating $T^{\mu\nu}$ in (1).  Since $T^{\mu\nu}$ is a regular tensor by assumption, the above steps all apply: the relevant, GR, field equations now include the matter stress-tensor as a right-hand side, while the Cotton sector stays source-free. This would seem to reduce everything to GR (now with a source) again; however, the Cotton sector, even if free, constrains its solutions. To study this, we use the ``kinematical" lemma of [2]: the only two non-identically vanishing components of $C^{\mu\nu}$ are those with one $C$-index along $X$ (here $\phi$), and its other either of the orthogonal $(r, t)$; these are also the only identically vanishing components of $G^{\mu\nu}$. From the definition (1),
\begin{eqnarray}
C^{r \phi} &=& \eps^{r t \phi}\,  [D_t\,  S_\phi ^\phi - D_\phi \, S_t^\phi] =0, \nonumber \\
C^{t \phi} &=& \eps^{t r \phi}\,  [D_r \, S_\phi ^\phi - D_\phi\,  S_r^\phi] =0.                     
\end{eqnarray}
What constraint, if any, does (4) impose on the Einstein-matter solutions? By circular symmetry, only $(T_{rr},T_{00},T_{0r})\ne 0$. Since we have restricted matter 
to $G_{\mu\nu}= \kappa \, T_{\mu\nu}$, and none of the corresponding three $G_{\mu\nu}$ components appears in (4), just the scalar curvature parts of $S_{\mu\nu}$ survive, and are manifestly required to have vanishing $r$- and $t$-derivatives ($\Lambda$-terms, being constant, never contribute in $C^{\mu\nu}$). But since this constant $R\sim T_\mu^\mu$, it vanishes for finite sources: we conclude that decoupling is permitted (only) in presence of null matter. This is not surprising physically, being driven by the remaining field equation, vanishing of the source-free Cotton-Weyl tensor. This is then a possibly interesting new class of explicit solutions of TMG. Note incidentally, that the converse type of source, pure ÒspinningÓ matter proportional to $\eps^{ij}$, hence coupled just to $C^{\mu\nu}$, is forbidden: the, now source-free, Einstein sector would require space to be (locally) flat.

 Some final remarks: First, our demonstration has not used the $X_\mu$-parallel/orthogonal component projection method of [2] explicitly; the familiar shortcuts afforded by circular symmetry clearly sufficed. Of course the two approaches fully agree as to which components vanish identically. Second, it should be clear that, by simply adapting coordinates, our construction fits any other HSOK. Third, note that some ``HSO"-aspects of $X_\mu$ were indeed essential: for example, Kerr-like solutions with non-HSO $X_\mu$ (involving essentially an explicit epsilon factor $\sim \eps^{ij}\,  x^j$ in the metric) do not decouple. The basic, metric tensor, variables must possess the HSOK symmetry; the only ``pseudo-"source is the explicit epsilon in Cotton. Given the latter's identical tracelessness, conformal HSOK might conceivably also suffice, but it seems unlikely that any other broad decoupling mechanisms exist.

We thanks S Carlip for reminding us of [2]. The work of SD was supported by grants NSF PHY 07-57190 and DOE DE-FG02-164 92ER40701.

\end{document}